\documentclass[hyper]{JHEP} 

\usepackage{epsfig}

\newcommand\fverb{\setbox\pippobox=\hbox\bgroup\verb}

\newcommand\fverbdo{\egroup\medskip\noindent%

            \fbox{\unhbox\pippobox}\ }

\newcommand\fverbit{\egroup\item[\fbox{\unhbox\pippobox}]}

\newbox\pippobox

\title{Reduced Sigma-Model on $O(N)$:
 Hamiltonian Analysis and  Poisson Bracket
of Lax Connection}

\author{J. Kluso\v{n}
 \footnote{On leave from Masaryk University, Brno}\\
Dipartimento di Fisica \& Sezione I.N.F.N.\\
Universit\`a di Roma
``Tor Vergata'' \\
Via della Ricerca, Scientifica 1 00133  Roma   ITALY\\
E-mail:
\email{Josef.Kluson@roma2.infn.it}}
\abstract{This short note is devoted to the
study of the Hamiltonian formalism and the 
integrability of  the bosonic model
introduced in [hep-th/0612079]. We
calculate Poisson bracket of spatial components
of Lax connection and we argue that its structure
implies  classical integrability of the theory.}


\def\tr{\mathrm{Tr}}

\def\mT{\mathcal{T}}

\def\pb  #1{\left\{#1\right\}}

\newcommand{\bA}{{\bf A}}
\newcommand{\bB}{{\bf B}}
\newcommand{\bC}{{\bf C}}

\newcommand{\mH}{\mathcal{H}}

\newcommand{\mA}{\mathcal{A}}

\newcommand{\mL}{\mathcal{L}}

\newcommand{\tsigma}{\tilde{\sigma}}

\begin{document}
\section{Introduction and Summary}
The sigma model  on $AdS_5\times
S^5$ \cite{Metsaev:1998it}
is complicated interacting theory
whose solution is currently beyond
the reach
\footnote{For alternative pure-spinor
approach description of superstring
theory on $AdS_5\times S^5$, see
\cite{Mikhailov:2007mr,Berkovits:2007zk,Berkovits:2000fe,
Berkovits:2000yr,Berkovits:2004xu,Vallilo:2002mh,Vallilo:2003nx,
Berkovits:2004jw,Bianchi:2006im,Kluson:2006wq}.}. 
On the other hand recently J. Maldacena and I. Swanson 
proposed a relatively simple kinematical truncation
of this theory \cite{Maldacena:2006rv}.
Technical simplifications allow us to test some
conjectures considering integrability of the
string theory on $AdS_5\times S^5$
\footnote{For some related works, see
\cite{Puletti:2007hq,Benvenuti:2007qt,
Das:2007tb,Chen:2007vs,Klose:2007rz,
Roiban:2007jf,Dorey:2007xn,Klose:2007wq}.}.
More precisely, the sigma model in this limit
leads  (after gauge fixing) to simpler 
toy model that is not, however, Lorentz-invariant
theory in $1+1$ dimensions.  On the other
hand it is well defined system on its own.
In fact, this system was carefully analysed
in \cite{Maldacena:2006rv} where world-sheet
S-matrix  was also discussed. Moreover, 
it was demonstrated an classical integrability
of the $O(N)$ sigma model in the near flat
space limit in the sense that the Lax pair
was constructed. This is very interesting
result since now we have Lax pair for completely
gauge fixed theory where the Virasoro constraints
were solved.  Since the
form of the Lax pair is rather unusual it is
interesting to study the property of this
theory further. 

It is well known that the existence of Lax
connection implies the existence of an infinite
tower of conserved charges in the classical
theory. However, as was stressed recently in
\cite{Dorey:2006mx} this does not quite
coincide with the standard definition of
integrability. Integrability in the standard
sense requires not only the existence of
a tower of conserved charges but also requires
that these charges are in "involution". In other
words, these conserved charges Poisson brackets
commute with each other. 

However there is a long-standing problem 
in determining the Poisson brackets
of the conserved charged for classical
string theory formulated on background
that admits Lax connection. Namely, the problem
is due to the presence of 
\emph{Non-Ultra Local} terms in the Poisson
brackets of the world-sheet fields that
lead to ambiguities in brackets for the charges
\footnote{For recent discussion of these
problems in the context of string theory
on $AdS_5\times S^5$, see 
\cite{Das:2007tb,Das:2005hp,Das:2004hy}.}.
As was shown recently in \cite{Dorey:2006mx}
a resolution of this problem is based on
earlier work of Maillet 
\cite{Maillet:1985ec,Maillet:1985ek} where
he proposed the regularisation the problematic
brackets. Then this procedure
was applied to the simplest classical
subsector of the $AdS_5\times S^5$
geometry in  \cite{Dorey:2006mx}. It
was shown that this prescription leads
to a very natural symplectic structure
on the space of finite-gap solutions 
of the string equations of motion that
were constructed in \cite{Dorey:2006zj}.
Then it was shown in 
\cite{Kluson:2007vw} that the string
theory on $AdS_5\times S^5$ possesses
 infinite number of conserved
charges that are in involution 
even on the world-sheet with general metric.
Explicitly, the Poisson bracket of spatial
components of Lax connection was calculated  
and it was shown that its form does not
depend on world-sheet metric and takes
precisely the form as in  \cite{Dorey:2006zj}.
However it is important to stress that
this analysis was valid in case 
when either the diffeomorphism invariance of theory
was preserved or the metric components
were fixed in some general form while 
the symmetry generated by Virasoro constraints
was preserved. On the other hand
it is not  clear whether integrability
is preserved in   case of completely fixed theory,
as for example, string theory in
 uniform light-cone gauge
\cite{Arutyunov:2004yx,Arutyunov:2006gs,Arutyunov:2005hd}
\footnote{This gauge was also discussed in
\cite{Astolfi:2007uz,Kluson:2007md}.}. 
As the first step in the
answering of this question we would like 
to calculate the Poisson bracket of the spatial
components of Lax connection for simpler model
introduced in   \cite{Maldacena:2006rv}. It turns
out that even if the resulting Poisson
bracket is very complicated one can map it, 
following \cite{Maillet:1985ec,Maillet:1985ek}
to the form that shows that the theory 
possesses an infinite number of local
charges that are in involutions. We mean
that this is a nice result that shows that
the integrability  persists in case of 
complete fixed theory as well. 

We can extend this work in various ways. 
For example, it would be nice to perform
the same analysis for the supersymmetric form
of the model given in \cite{Maldacena:2006rv}.
Then  we would like to  calculate 
Poisson brackets of Lax connection
for bosonic string in uniform light-cone gauge. 

The organisation of this paper is as follows.
In next section (\ref{second}) we review
the procedure presented in \cite{Maldacena:2006rv} for
$O(N)$ sigma model that leads to simpler 
model with completely fixed symmetry.
In section (\ref{third}) we review the construction 
of Lax connection for this system.
Then in section (\ref{fourth}) we present
the Hamiltonian formalism of this system
and we calculate the Poisson brackets of
spatial components of Lax connection.
Using these results we argue  for
integrability of the theory. Finally, in
Appendix (\ref{fifth}) we review 
Maillet's treatment of the monodromy 
matrix and Lax connection.
\section{Reduced  
$O(N)$ Sigma Model}\label{second}
In this section we review the
analysis presented in \cite{Maldacena:2006rv}
that leads to interesting new $1+1$
dimensional field theory. We consider
$O(N)$ sigma model. The target space
of this sigma model is a sphere $S^{N-1}$.
Let us consider a state with a constant
spin density $J=J_{12}$, where 
$J_{kl}$ are rotation generators
in the $kl$ plane. 
Then we  begin with the action
\begin{eqnarray}\label{actO}
S&=&-\frac{\sqrt{\lambda}}{4\pi}
\int d\tsigma^0 d\tsigma^1
\sqrt{-\eta}(-\eta^{\alpha\beta}
\partial_\alpha t\partial_\beta t+\eta^{\alpha\beta}
\partial_\alpha x^m\partial_\beta x^n) \ , \nonumber \\
& & 
 x^mx^n\delta_{mn}=1 \ , \quad 
 m= 1,\dots,N \ , 
\nonumber \\
\end{eqnarray}
where $\partial_\alpha\equiv 
\frac{\partial}{\partial \tsigma^\alpha} \ , 
\alpha=0,1$ and we work in conformal gauge 
with the world-sheet metric $\eta_{\alpha\beta}=
\mathrm{diag}(-1,1)$. To have a contact
with  \cite{Maldacena:2006rv}
we introduce the parameter
\begin{equation}\label{defg}
g=\frac{\sqrt{\lambda}}{4\pi} \ . 
\end{equation}

For simplicity, we start with 
the case when 
$N=3$ and parametrize  $S^2$
as
\begin{equation}
x^1=\sin\theta
\ , \quad x^2=\cos\theta \cos\phi \ , 
\quad x^3=\cos\theta \sin \phi \ .
\end{equation}
Then the action (\ref{actO}) 
 takes the form 
\begin{equation}\label{act1a}
S=2g\int d\tsigma^+ d\tsigma^-
(-\partial_{\tsigma^+} t
\partial_{\tsigma^-} t+
\cos^2\theta \partial_{\tsigma^+}\phi
\partial_{\tsigma^-}\phi+
\partial_{\tsigma^+}
\theta\partial_{\tsigma^-}\theta) \ ,
\end{equation}
where we have introduced 
the light-cone coordinates 
\footnote{In the light-cone frame
the metric components are 
$\eta^{+-}=\eta^{-+}=-2$, with the inverse
$\eta_{+-}=\eta_{-+}=-\frac{1}{2}$ and
with the corresponding determinant $
\sqrt{-\eta}=\frac{1}{2}$.}
\begin{equation}
\tsigma^\pm=\tsigma^0\pm \tsigma^1 
\ .
\end{equation}
Further, since the action (\ref{act1a})
is defined with fixed form of the
world-sheet metric the theory has
to be accompanied with 
the corresponding  Virasoro constrains 
\begin{eqnarray}\label{Vircor}
T_{++}&=&
g[-\partial_{\tsigma^+}t\partial_{\tsigma^+}t+
\cos^2\theta \partial_{\tsigma^+}
\phi\partial_{\tsigma^+}\phi+
\partial_{\tsigma^+}\theta\partial_{\tsigma^+}
\theta]=0 \ , 
\nonumber \\
T_{--}&=&
g[-\partial_{\tsigma^-}t\partial_{\tsigma^-}t+
\cos^2\theta \partial_{\tsigma^-}
\phi\partial_{\tsigma^-}\phi+
\partial_{\tsigma^-}\theta\partial_{\tsigma^-}
\theta]=0 \ .
\nonumber \\
\end{eqnarray}
Our goal is to consider state
with a constant
spin density $J_{12}=J$. Let us
start with solution: $\frac{d\phi}
{d\tsigma^0}=0$  and
$\theta=0$ and perform a boost on the
world-sheet coordinates $\tsigma^\pm=
\tsigma^0\pm \tsigma^1$ and expand in small
fluctuations around the constant spin-density
solution
\begin{eqnarray}\label{rescall}
\tsigma^+&=&2\sqrt{g}
\sigma^+ \ , \quad \tsigma^-=
\frac{\sigma^-}{2\sqrt{g}} \ , 
\nonumber \\
\phi&=&\tsigma^0+\frac{\delta}{\sqrt{g}}=
\frac{\tsigma^++\tsigma^-}{2}+
\frac{\delta}{\sqrt{g}}=\nonumber \\
&=&\sqrt{g}\sigma^++\frac{\chi }{\sqrt{g}} \ , \quad 
\chi=\frac{\sigma^-}{4}+\delta \ , 
\nonumber \\
\theta&=&\frac{y}{\sqrt{g}} \ , 
 \quad \nonumber \\
t&=&\tsigma^0=\frac{1}{2}
(\tsigma^++\tsigma^-)=
\sigma^++\frac{\sigma^-}{4\sqrt{g}}
\ ,  \quad 
g\rightarrow \infty \ ,  \nonumber \\
\end{eqnarray}
where $\sigma^\pm$ are the light-cone 
coordinates
after performing the boost. Note 
that we are interested in solutions
where $\chi=\frac{1}{4}\sigma^-+\delta$ 
with $\delta$ representing  small 
fluctuations.
Then the rescaling (\ref{rescall})
 implies
\begin{eqnarray}\label{rescall1}
d\tsigma^+d\tsigma^-&=&
d\sigma^+ d\sigma^- \ , \nonumber \\ 
\frac{\partial}{\partial
\tsigma^+}&=&
\frac{1}{2\sqrt{g}}\frac{\partial}
{\partial \sigma^+}\equiv
\frac{1}{2\sqrt{g}}\partial_+ \ , 
 \nonumber \\
 \frac{\partial}{\partial
 \tsigma^-}&=&2\sqrt{g}\frac{\partial}
 {\partial \sigma^-}\equiv
 2\sqrt{g}\partial_- \ .  \nonumber \\
\end{eqnarray}
Using (\ref{rescall}) and (\ref{rescall1})
the  action (\ref{act1a}), up to constant
and total derivative terms, is finite
\begin{eqnarray}
S=
2\int d\sigma^+ d\sigma^-
[\partial_+\chi\partial_-\chi+
\partial_+y\partial_-y-y^2
\partial_-\chi] \ .  \nonumber \\
\end{eqnarray}
Finally, to generalize to the case
of $O(N)$ 
sigma model we replace $y^2$ with
$y^2\rightarrow \vec{y}^2\equiv y^iy_i \ ,
i=1,\dots,N-2$
in the above action and we obtain
\begin{equation}\label{resact}
S=2\int d\sigma^+ d\sigma^-
[\partial_-\chi\partial_+\chi
+\partial_+\vec{y}\partial_-\vec{y}
-\vec{y}^2\partial_-\chi] \ . 
\end{equation}
Note also that under the
rescaling (\ref{rescall}), 
$T_{++}$ given in (\ref{Vircor})
 takes the form
\begin{eqnarray}
\lim_{g\rightarrow \infty}
T_{++}
=\frac{1}{2}\left(
\partial_+\chi-\frac{\vec{y}^2}{2}\right)+
O(\frac{1}{g})
\equiv
\frac{1}{2}j_++
O(\frac{1}{g})
 \nonumber \\
\end{eqnarray}
and hence the 
 Virasoro constraint $T_{++}=0$
 implies the constraint 
\begin{equation}
j_+\equiv\partial_+\chi-\frac{\vec{y}^2}{2}=0 \ . 
\end{equation}
 In the
same way the rescaling (\ref{rescall})
performed on $T_{--}$ implies 
\begin{eqnarray}
\partial_-\chi\partial_-\chi+\partial_-\vec{y}
\partial_-\vec{y}=16 \ . \nonumber \\
\end{eqnarray}
In summary, we have two constraints
in the theory
\begin{eqnarray}\label{gfc}
\Phi_1&=& \partial_-\chi\partial_-\chi+\partial_-\vec{y}
\partial_-\vec{y}-16=0 \ , \nonumber \\
\Phi_2&=&\partial_+\chi-\frac{\vec{y}^2}{2}=0 \ . 
\nonumber \\
\end{eqnarray}
Following  \cite{Maldacena:2006rv}
 we can move to a gauge fixed Lagrangian
by defining new coordinates
\begin{equation}\label{defxplus}
x^+\equiv  \sigma^+ \ ,
\quad x^-\equiv \frac{\sigma^-}{2}+2\chi \ . 
\end{equation}
Then it is easy to see that
\begin{eqnarray}\label{defxplusd}
\partial_{+}&=&
\partial_{x^+}+2\partial_+\chi
\partial_{x^-}=
\partial_{x^+}+\vec{y}^2\partial_{x^-} \ , 
\nonumber \\
\partial_-&=&
2(\frac{1}{4}+\partial_-\chi)
\partial_{x^-} \ , \nonumber \\
\end{eqnarray}
where we have used  (\ref{gfc}).
Note also that 
$\Phi_2$ implies
\begin{eqnarray}\label{defxplusdc}
\partial_-\chi=
\frac{\frac{1}{4}-(\partial_{x^-}\vec{y})^2}
{1+4(\partial_{x^-}\vec{y})^2} \ , \quad 
\frac{\partial_-\chi}{
\frac{1}{4}+\partial_-\chi}=
2(
\frac{1}{4}-(\partial_{x^-}\vec{y})^2) \ . 
\nonumber \\
\end{eqnarray}
Let us now consider the equations of
motion for $y^i$ that follow from
the action (\ref{resact})
\begin{equation}\label{eqyi}
\partial_{+}\partial_{-}
y^i+y^i\partial_{-}\chi=0 \ .
\end{equation}
Then using (\ref{defxplusd}) and
 (\ref{defxplusdc}) we 
can map (\ref{eqyi}) into 
\begin{eqnarray}
\partial_{x^-}\partial_{x^+}y^i+
\partial_{x^-}(\vec{y}^2\partial_{x^-}y^i)+
y^i(
\frac{1}{4}-(\partial_{x^-}y)^2)=0 \ .  \nonumber \\
\end{eqnarray}
Then it is easy to see that 
 these equations of motion follow
from the variation of the action
\begin{equation}\label{actH}
S=2\int dx^+dx^-
[\partial_{x^+}\vec{y}\partial_{x^-}\vec{y}-\frac{1}{4}
\vec{y}^2+\vec{y}^2(\partial_{x^-}\vec{y}\partial_{x^-}\vec{y})] \ .
\end{equation}
The action (\ref{actH}) will be starting point
for the Hamiltonian treatment of the reduced
model. Before we proceed to this question we
introduce the Lax connection for the
theory given above. 
\section{Lax connection}\label{third}
In this section we introduce Lax connection
that was given in \cite{Maldacena:2006rv}.
As was argued there Lax
connection can be obtained by taking
a simple limit of the connection
of $O(N)$ theory. In order to write
it explicitly, we select one
of the $O(N)$ generators $J_{12}$
and consider the off-diagonal generators
that mix $(1,2)$ plane with the rest. 
We denote these generators as $J^{\pm i}\ ,
i=1,\dots,N-2$. We also need following
commutation relations:
\begin{eqnarray}\label{comJ}
\left[J^{12},J^{\pm i}\right]&=&\pm J^{\pm i } \ , 
\nonumber \\
\left[J^{+ i},J^{-j}\right]&=&\delta_{ij}J^{12}
-J^{ij} \ , 
\quad
\left[J^{-i},J^{+j}\right]=-\delta_{ij}J^{12}
-J^{ij} \ . \nonumber \\
\end{eqnarray}
Flat connection introduced  in 
\cite{Maldacena:2006rv} is derived
by taking the limit of the connection
for $O(N)$ theory \cite{Beisert:2004ag}
and it takes the form
\begin{eqnarray}\label{mAm}
\mA_+&=&\frac{i}{\sqrt{2}}
\left[e^{-i\sigma^+ w} y^i J^{+i}
+e^{i\sigma^+w}y^iJ^{-i}\right] \ ,
\nonumber \\
\mA_-&=&
\frac{1}{w}
\left[-i\partial_-\chi J^{12}
-\frac{1}{\sqrt{2}}
e^{-i\sigma^+w}\partial_-y^i J^{+i}
+\frac{1}{\sqrt{2}}
e^{i\sigma^+ w}\partial_-y^i J^{-i}
\right] \ ,
\nonumber \\
\end{eqnarray}
where $w$ is spectral parameter. 
Then using the constraint $j_+=0$ 
and  the equations of motion for $y^i$
(\ref{eqyi})
together with 
(\ref{comJ})
we obtain 
\begin{equation}
\partial_+\mA_--\partial_-\mA_++
[\mA_+,\mA_-]=0 \ . 
\end{equation}
In other words $\mA$ given
(\ref{mAm}) defines 
flat Lax connection.
As the next step we use 
(\ref{defxplus}) and write
\begin{eqnarray}
& &dx^+= d\sigma^+ \ , 
\nonumber \\ 
& & dx^--\vec{y}^2 dx^+=d\sigma^-2
(\frac{1}{4}+\partial_{-}\chi) \ . 
\nonumber \\
\end{eqnarray}
Then we obtain
\begin{eqnarray}
\mA=\mA_+d\sigma^++\mA_- d\sigma^-
=\mA_+dx^++\tilde{\mA}
(dx^--\vec{y}^2 dx^+) \ , 
\nonumber \\
\end{eqnarray}
where we have defined
\begin{eqnarray}
\tilde{\mA}&\equiv& 
\mA_-\frac{1}{2(\frac{1}{4}+
\partial_-\chi)}=\nonumber \\
&=&\frac{1}{w}
\left[-i(\frac{1}{4}-(\partial_{x^-}
\vec{y})^2)J^{12}-\frac{1}{\sqrt{2}}
e^{-ix^+w}\partial_{x^-}y^i
J^{+i}+
\frac{1}{\sqrt{2}}e^{ix^+w}\partial_{x^-}
y^i J^{-i}\right] \ , 
\nonumber \\
\end{eqnarray}
where in the final step we used
 (\ref{defxplusd}) and (\ref{defxplusdc}).
Finally, we can perform gauge transformation
to remove the constant part of the connection 
\begin{equation}
\mA'=g^{-1}\mA g+g^{-1}dg \ , \quad 
g=e^{i\frac{1}{4w}x^- J^{12}}
\end{equation}
and we obtain
\begin{eqnarray}\label{mA'}
\mA_+'&=&\frac{i}{\sqrt{2}}
\left[e^{-ix^+w-\frac{i}{4w}x^-}y^i J^{+i}+
e^{ix^+w+i\frac{1}{4w}x^-}y^i J^{-i}\right]
\ , \nonumber \\
\tilde{\mA}'&=&
\frac{1}{w}
\left[i(\partial_{x^-}y)^2J^{12}-
\frac{1}{\sqrt{2}}e^{-ix^+w-i\frac{1}{4w}
x^-}\partial_{x^-}y^iJ^{+i}+\right. \nonumber \\
& & \left.+\frac{1}{\sqrt{2}}e^{ix^+w+i\frac{1}{4w}
x^-}\partial_{x^-}y^i J^{-i}\right] \ . 
\nonumber \\
\end{eqnarray}
In what follows we use the Lax
connection (\ref{mA'}) where
we will write $\mA$ instead of
$\mA'$.  
Let us now rewrite the Lax connection as
\begin{eqnarray}
\mA&=&(\mA_+-\vec{y}^2\tilde{\mA}_-)dx^++
\tilde{\mA}_-dx^-=
\nonumber \\
&=&(\mA_+-\vec{y}^2\tilde{\mA}_-+\tilde{\mA}_-)
d\tau+(\mA_+-\vec{y}^2\tilde{\mA}_--\tilde{\mA}_-)
d\sigma  \ , 
\end{eqnarray}
where we have introduced $\sigma,\tau$ defined
as 
\begin{equation}
x^\pm=\tau\pm \sigma  \ . 
\end{equation}
We see that the 
 spatial component of Lax connection 
$\mA_\sigma$ takes the form 
\begin{eqnarray}\label{mAsigma}
\mA_\sigma&=&
(\mA_+-\vec{y}^2\tilde{\mA}_--\tilde{\mA}_-)=
\nonumber \\
&=&\frac{i}{\sqrt{2}}
\left[e^{-ix^+w-\frac{i}{4w}x^-}y^i J^{+i}+
e^{ix^+w+i\frac{1}{4w}x^-}y^i J^{-i}\right]-
\nonumber \\
&-&(1+\vec{y}^2)
\frac{1}{w}
\left[i(\partial_{x^-}y)^2J^{12}-
\frac{1}{\sqrt{2}}e^{-ix^+w-i\frac{1}{4w}
x^-}\partial_{x^-}y^iJ^{+i}+\right. \nonumber \\
& &\left.+\frac{1}{\sqrt{2}}e^{ix^+w+i\frac{1}{4w}
x^-}\partial_{x^-}y^i J^{-i}\right] \ . 
\nonumber \\
\end{eqnarray}
The spatial component of Lax connection given
above will be the central object for the study 
of the integrability of the theory. Explicitly,
we will calculate the Poisson bracket between
these components for different spectral parameters
$w,v$. Before we proceed to this calculation we
have to develop corresponding Hamiltonian formalism.
\section{Hamiltonian formalism}\label{fourth}
Our goal is to develop the Hamiltonian
formalism for the action
\begin{equation}
S=4\int dx^+dx^- \sqrt{-\eta}
[\partial_{x^+}\vec{y}\partial_{x^-}\vec{y}-\frac{1}{4}
\vec{y}^2+\vec{y}^2(\partial_{x^-}\vec{y}
\partial_{x^-}\vec{y})] \ . 
\end{equation}
If we again introduce
 coordinates $\tau,\sigma$
as
\begin{equation}
x^\pm=\tau\pm \sigma 
\end{equation}
and consequently 
\begin{equation}
\partial_{x^+}=
\frac{1}{2}(\partial_\tau+\partial_\sigma) \ ,
\quad
\partial_{x^-}=
\frac{1}{2}(\partial_\tau-\partial_\sigma)
\end{equation}
we obtain
\begin{equation}
S=
\int d\tau d\sigma
[(\partial_\tau \vec{y})^2-(\partial_\sigma
\vec{y})^2-\vec{y}^2+\vec{y}^2(\partial_\tau \vec{y}-
\partial_\sigma \vec{y})^2] \ . 
\end{equation}
Then the momentum $\pi_i$ conjugate to $y^i$ 
takes the form
\begin{equation}\label{pii}
\pi_i=\frac{\delta L}{\delta\partial_\tau
y^i}=
2[\partial_\tau y^i+\vec{y}^2(\partial_\tau y^i
-\partial_\sigma y^i)]
\end{equation}
and we have following 
 canonical Poisson brackets 
\begin{equation}
\pb{y^i(\sigma),\pi_j(\sigma')}=
\delta^i_j\delta(\sigma-\sigma') \ . 
\end{equation}
Using (\ref{pii}) we can express
$\partial_\tau y^i$ as function
of $\pi_i$ and $\partial_\sigma y^i$
\begin{equation}\label{partauy}
\partial_\tau y^i=
\frac{\frac{1}{2}\pi_i+\vec{y}^2\partial_\sigma y^i}
{1+\vec{y}^2} \ . 
\end{equation}
Then corresponding Hamiltonian density
takes the form
\begin{eqnarray}
\mH&=&\partial_\tau y^i\pi_i-\mL=\nonumber \\
&=&\frac{\vec{\pi}^2}{4(1+\vec{y}^2)}+\frac{\vec{y}^2
(\partial_\sigma \vec{y}\vec{\pi})}{1+\vec{y}^2}
-\frac{\vec{y}^2(\partial_\sigma \vec{y})^2}
{1+\vec{y}^2}+(\partial_\sigma \vec{y})^2+
\vec{y}^2 \ . 
\nonumber \\
\end{eqnarray}
Finally, we use the relation 
(\ref{partauy}) to express 
(\ref{mAsigma})
as a function of canonical
variables
\begin{eqnarray}
\mA_\sigma
&=&\frac{i}{\sqrt{2}}
\left[e^{-ix^+w-\frac{i}{4w}x^-}y^i J^{+i}+
e^{ix^+w+i\frac{1}{4w}x^-}y^i J^{-i}\right]-
\nonumber \\
&-&\frac{1}{w}
\left[\frac{i}{4(1+\vec{y}^2)}
(\frac{1}{2}\vec{\pi}-\partial_\sigma \vec{y})^2J^{12}-
\frac{1}{2\sqrt{2}}e^{-ix^+w-i\frac{1}{4w}
x^-}(\frac{1}{2}\pi_i-\partial_\sigma y^i)
J^{+i}+\right. \nonumber \\
& &\left.+\frac{1}{2\sqrt{2}}e^{ix^+w+i\frac{1}{4w}
x^-}(\frac{1}{2}\pi_i-\partial_\sigma y^i) J^{-i}\right] \ .
\nonumber \\
\end{eqnarray}
Now we are ready to calculate the Poisson
brackets 
\begin{equation}
\pb{\mA_{\sigma,\alpha\beta}(\sigma,w),
\mA_{\sigma,\gamma\delta}(\sigma',v)} \ ,  
\end{equation}
where  $\alpha,\beta$ and $\gamma,\delta$ 
label  matrix indices of generators $J$'s.
 Since
in the following we will consider the spatial
components of Lax connection only we omit
the subscript $\sigma$. Explicitly we define
$\mA_{\sigma,\alpha\beta}\equiv
\mA_{\alpha\beta}$. 

Now using the canonical Poisson brackets
determined above we calculate the Poisson
bracket of spatial component of Lax connection.
Using 
\begin{eqnarray}
\pb{(\frac{1}{2}\vec{\pi}-\partial_\sigma \vec{y})^2(\sigma),
(\frac{1}{2}\pi_i-\partial_{\sigma'}y^i)(\sigma')}&=&
(\frac{1}{2}\pi-\partial_\sigma y^i)(\sigma)[\partial_{\sigma'}
\delta(\sigma-\sigma')
-\partial_\sigma \delta(\sigma-\sigma')]  \ , 
\nonumber \\
\pb{(\frac{1}{2}\pi_i-\partial_\sigma y^i)(\sigma),
(\frac{1}{2}\vec{\pi}-\partial_{\sigma'}\vec{y})^2(\sigma')}&=&
-(\frac{1}{2}\pi_i-\partial_{\sigma'} y^i)(\sigma')
[\partial_\sigma \delta(\sigma-\sigma')-
\partial_{\sigma'}\delta(\sigma-\sigma')] \ , 
\nonumber \\
\pb{(\frac{1}{2}\vec{\pi}-\partial_\sigma \vec{y})^2(\sigma),
(\frac{1}{2}\vec{\pi}-\partial_{\sigma'}\vec{y})^2(\sigma')}&=&
2(\frac{1}{2}\pi_i-\partial_\sigma y^i)(\sigma)
[\partial_{\sigma'}
\delta(\sigma-\sigma')-\nonumber \\
&-&\partial_\sigma
\delta(\sigma-\sigma')]
(\frac{1}{2}\pi_i-\partial_{\sigma'}y^i)(\sigma')
\nonumber \\
\end{eqnarray}
we obtain, after  straightforward
calculations, following result 
\begin{eqnarray}\label{pbmA1}
& &\pb{\mA_{\alpha\beta}(\sigma,w),
\mA_{\gamma\delta}(\sigma',v)}=
\bA_{\alpha\gamma,\beta\delta}(\sigma,w,v)
\delta(\sigma-\sigma')+\nonumber \\
&+&\bB_{\alpha\gamma,\beta\delta}(\sigma,\sigma',w,v)
\partial_{\sigma'} \delta(\sigma-\sigma')+
\bC_{\alpha\gamma,\beta\delta}(\sigma,\sigma',w,v)
\partial_{\sigma}\delta(\sigma-\sigma') \ ,
\nonumber \\
\end{eqnarray}
where 
\begin{eqnarray}
& &\bA_{\alpha\gamma,\beta\delta}(\sigma,w,v)=
\nonumber \\
&=&
\frac{i(w-v)}{8vw}
e^{-ix^+(v+w)-\frac{i(v+w)}{4vw}x^-}
J^{+i}_{\alpha\beta}J^{+i}_{\gamma\delta}-
\frac{i(v+w)}{8vw}
e^{-ix^+(w-v)-\frac{i(v-w)}{4vw}x^-}
J^{+i}_{\alpha\beta}J^{-i}_{\gamma\delta}+
\nonumber \\
&+&\frac{i(v+w)}{8vw}
e^{i(w-v)x^++\frac{i(v-w)}{4wv}x^-}
J^{-i}_{\alpha\beta}J^{+i}_{\gamma\delta}-
\frac{i(w-v)}{8wv}
e^{ix^+(v+w)+\frac{i(v+w)}{4vw}x^-}
J^{-i}_{\alpha\beta}J^{-i}_{\gamma\delta}+
\nonumber \\
&+&\frac{i}{4w\sqrt{2}
(1+\vec{y}^2)}[(\frac{1}{2}\pi_i-\partial_\sigma y^i)
e^{-ix^+v-\frac{i}{4v}x^-}J^{12}_{\alpha\beta}
J^{+i}_{\gamma\delta}+
e^{ix^+v +\frac{i}{4v}x^-}(\frac{1}{2}\pi_i-\partial_\sigma y^i)
J^{12}_{\alpha\beta}
J^{-i}_{\gamma\delta}]
-\nonumber \\
&-&\frac{i}{4v\sqrt{2}(1+\vec{y}^2)}
[(\frac{1}{2}
\pi_i-\partial_\sigma y^i)e^{-ix^+w-\frac{i}{4w}x^-}
J^{+i}_{\alpha\beta}J^{12}_{\gamma\delta}
+e^{ix^+w+\frac{i}{4w}x^-}(\frac{1}{2}\pi_i-\partial_\sigma y^i)
J^{-i}_{\alpha\beta}J_{\gamma\delta}^{12}]+
\nonumber \\
&+&\frac{i}{16 vw}
\frac{1}{(1+\vec{y}^2)^2}
(\frac{1}{2}\vec{\pi}-\partial_\sigma \vec{y})^2
[ e^{-ix^+v-\frac{i}{4v}x^-}y^i
J^{12}_{\alpha\beta}
J^{+i}_{\gamma\delta}
-e^{ix^+v+\frac{i}{4v}x^-}y^i
J^{12}_{\alpha\beta}
J^{-i}_{\gamma\delta}
-\nonumber \\
&-&e^{-ix^+w-\frac{i}{4w}x^-}y^iJ^{+i}_{\alpha\beta}
J^{12}_{\gamma\delta}+
e^{ix^+w+\frac{i}{4w}x^-}y^iJ^{-i}_{\alpha\beta}
J^{12}_{\gamma\delta}] \ , 
\nonumber \\
\end{eqnarray}
\begin{eqnarray}
& &\bC_{\alpha\gamma,\beta\delta}(\sigma,\sigma',
w,v)=\nonumber \\
&=&
-\frac{1}{16 vw}
[e^{-ix^+w-\frac{i}{4w}x^-}J^{+i}_{\alpha\beta}
-e^{ix^+w+\frac{i}{4w}x^-}J^{-i}_{\alpha\beta}](\sigma)\times
\nonumber \\
&\times& [e^{-ix^+v-\frac{i}{4v}x^-}J^{+i}_{\gamma\delta}
-e^{ix^+v+\frac{i}{4v}x^-}J^{-i}_{\gamma\delta}](\sigma')+
\nonumber \\
&+&\frac{i}{16 vw}
\frac{1}{(1+\vec{y}^2)}
(\frac{1}{2}\pi_i-\partial_\sigma y^i)(\sigma)
  [e^{-ix^+v-\frac{i}{4v}x^-}J^{12}_{\alpha\beta}
J^{+i}_{\gamma\delta}
-e^{ix^+v+\frac{i}{4v}x^-}J^{12}_{\alpha\beta}
J^{-i}_{\gamma\delta}]
(\sigma')+ \nonumber \\
&+&\frac{i}{16 v w}[e^{-ix^+w-\frac{i}{4w}x^-}J^{+i}_{\alpha\beta}
J^{12}_{\gamma\delta}
-e^{ix^+w+\frac{i}{4w}x^-}
J^{-i}_{\alpha\beta}J^{12}_{\gamma\delta}](\sigma)
\frac{(\frac{1}{2}\pi_i-
\partial_{\sigma'} y^i)}{(1+\vec{y}^2)}(\sigma')-
\nonumber \\
&-&\frac{1}{8 wv}
\frac{(\frac{1}{2}
\pi_i-\partial_\sigma y^i)}{1+\vec{y}^2}
(\sigma)\frac{(\frac{1}{2}\pi_i-\partial_{\sigma'}y^i)}
{1+\vec{y}^2}(\sigma')
J^{12}_{\alpha\beta}J^{12}_{\gamma\delta}
\nonumber \\
\end{eqnarray}
and 
\begin{eqnarray}
& &\bB_{\alpha\gamma,\beta\delta}(\sigma,\sigma',
w,v)= \nonumber \\
&=&\frac{1}{16 vw}
[e^{-ix^+w-\frac{i}{4w}x^-}J^{+i}_{\alpha\beta}
-e^{ix^+w+\frac{i}{4w}x^-}J^{-i}_{\alpha\beta}](\sigma)\times
\nonumber \\
&\times& [e^{-ix^+v-\frac{i}{4v}x^-}J^{+i}_{\gamma\delta}
-e^{ix^+v+\frac{i}{4v}x^-}J^{-i}_{\gamma\delta}](\sigma')-
\nonumber \\
&-&\frac{i}{16 vw}
\frac{1}{(1+\vec{y}^2)}
(\frac{1}{2}
\pi_i-\partial_\sigma y^i)(\sigma)
[e^{-ix^+v-\frac{i}{4v}x^-}
J^{12}_{\alpha\beta}J^{+i}_{\gamma\delta}
-e^{ix^+v+\frac{i}{4v}x^-}
J^{12}_{\alpha\beta}J^{-i}_{\gamma\delta}](\sigma')-
\nonumber \\
&-&\frac{i}{16 v w}[e^{-ix^+w-\frac{i}{4w}x^-}J^{+i}_{\alpha\beta}
J^{12}_{\gamma\delta}
-e^{ix^+w+\frac{i}{4w}x^-}J^{-i}_{\alpha\beta}J^{12}_{\gamma\delta}
](\sigma)
\frac{(\frac{1}{2}\pi_i-
\partial_{\sigma'} y^i)}{(1+\vec{y}^2)}(\sigma')+
\nonumber \\
&+&\frac{1}{8 wv}
\frac{(\frac{1}{2}
\pi_i-\partial_\sigma y^i)}{1+\vec{y}^2}
(\sigma)\frac{(\frac{1}{2}
\pi_i-\partial_{\sigma'}y^i)}
{1+\vec{y}^2}(\sigma')
J^{12}_{\alpha\beta}J^{12}_{\gamma\delta} \ . 
\nonumber \\
\end{eqnarray}
Then it is easy to see that matrices
$\bA,\bB,\bC$ obey the relations
\begin{eqnarray}
\bA_{\alpha\gamma,\beta\delta}(\sigma,
w,v)=
-\bA_{\gamma\alpha,
\delta\beta}(\sigma,v,w)
\nonumber \\
\end{eqnarray}
and 
\begin{eqnarray}
\bB_{\alpha\gamma,\beta\delta}(\sigma,\sigma',
w,v)=
-\bC_{\gamma\alpha,\beta\delta}(\sigma',\sigma,v,w) \ 
\nonumber \\
\end{eqnarray}
that  are in agreement 
with  general definition given 
in (\ref{gendefbA}).

Now we are ready to exhibit the general
structure of the Poisson brackets, following 
\cite{Maillet:1985ec}.
 Let us introduce the matrices
$r_{\alpha\gamma,\beta\delta}(\sigma,w,v),
s_{\alpha\gamma,\beta\delta}(\sigma,w,v)$ whose 
explicit form in terms of matrices
$\bB_{\alpha\gamma,\beta\delta},\bC_{\alpha\gamma,
\beta\delta}$ is given in (\ref{rsdef}). 
Then, using also 
the formula 
\begin{equation}
f(x,y)\partial_x\delta(x-y)=
f(x,x)\partial_x\delta(x-y)+
\partial_y f(x,y)_{y=x}\delta(x-y) \ 
\end{equation}
 we can rewrite
the Poisson bracket (\ref{pbmA1}) into
the form 
\begin{eqnarray}\label{pbmA1a}
& &\pb{\mA_{\alpha\beta}(\sigma,w),
\mA_{\gamma\delta}(\sigma',v)}=
\bA_{\alpha\gamma,\beta\delta}(\sigma,w,v)
\delta(\sigma-\sigma')-\nonumber \\
&-&\partial_{u}\bB_{\alpha\gamma,
\beta\delta}(\sigma,u,w,v)_{u=\sigma}
\delta(\sigma-\sigma')-
\partial_{u}\bC_{\alpha\gamma,\beta\delta}
(u,\sigma,w,v)_{u=\sigma}
\delta(\sigma-\sigma')-
\nonumber \\
&-&\bB_{\alpha\gamma,\beta\delta}(\sigma,\sigma,w,v)
\partial_{\sigma}\delta(\sigma-\sigma')
-
\bC_{\alpha\gamma,\beta\delta}(\sigma',\sigma',w,v))
\partial_{\sigma'}\delta(\sigma-\sigma')= 
 \nonumber \\
&=&(\partial_\sigma
r_{\alpha\gamma,\beta\delta}
(\sigma,w,v)-
\partial_\sigma s_{\alpha\gamma,\beta\delta}
(\sigma,w,v))\delta(\sigma-\sigma')
-2
s_{\alpha\gamma,\beta\delta}(\sigma,w,v)
\partial_{\sigma}\delta(\sigma-\sigma')
+\nonumber \\
&+&[(r_{\alpha\gamma,\sigma\delta}(\sigma,w,v)-
s_{\alpha\gamma,\sigma \delta}
(\sigma,w,v))
\mA_{\sigma\beta}(\sigma,w)-
 \nonumber \\&-&
\mA_{\alpha\sigma}(\sigma,w)
(r_{\sigma\gamma,\beta\delta}(\sigma,w,v)
-s_{\sigma\gamma,\beta\delta}
(\sigma,w,v))]\delta(\sigma-\sigma')+
\nonumber \\
&+&[(r_{\alpha\gamma,\beta\sigma}(\sigma,w,v)+
s_{\alpha\gamma,\beta\sigma}(\sigma,
w,v))
\mA_{\sigma\delta}(v,\sigma)-\nonumber \\
&-&\mA_{\gamma\sigma}(v,\sigma)(r_{\alpha\sigma,\beta\delta}
(\sigma,w,v)+s_{\alpha\sigma,\beta\delta}
(\sigma,w,v))]\delta(\sigma-\sigma') \ . 
\nonumber \\ 
\end{eqnarray}
The fact that the Poisson bracket
of Lax connection takes the form given
above has an important consequence for the
integrability of the theory. As was shown in 
\cite{Maillet:1985ec} and reviewed in
Appendix integrable theories with the
Poisson brackets of the Lax connection
given in (\ref{pbmA1a}) or in its
alternative form given in (\ref{pbMaillet}) possesses 
infinite number of conserved charges
that are in involution with respect to given
Poisson bracket structure. 
In other words we have shown that the
reduced sigma model is classically integrable. 
\section{Appendix: Review of Basic Properties
of Monodromy Matrix}\label{fifth}
In this section we give a review of 
properties of monodromy matrix, following 
\cite{Maillet:1985ek}. As opposite to this
paper we will write all expressions with
explicit matrix notation. 

The monodromy 
matrix $\mT_{\alpha\beta}(\sigma_1,\sigma_2,w)$, where
$w$ is a spectral parameter, can
be defined as 
\begin{eqnarray}\label{diffdefT}
\partial_{\sigma_1} \mT_{\alpha\beta}
(\sigma_1,\sigma_2,w)&=&
\mA_{\alpha\gamma}(\sigma_1,w)
\mT_{\gamma\beta}(\sigma_1,\sigma_2,w) \ ,
\nonumber \\
\partial_{\sigma_2}
 \mT_{\alpha\beta}(\sigma_1,\sigma_2,w)&=&
-\mT_{\alpha\gamma}
(\sigma_1,\sigma_2,w)
\mA_{\gamma\beta}(\sigma_2,w) \ 
\nonumber \\
\end{eqnarray}
with the normalisation condition
\begin{equation}
\mT_{\alpha\beta}
(\sigma_1,\sigma_2,w)=\delta_{\alpha\beta} \ 
\end{equation}
and  
\begin{equation}
\mT^{-1}_{\alpha\beta}(\sigma_1,\sigma_2,w)=
\mT_{\alpha\beta}(\sigma_2,\sigma_1,w) \ . 
\end{equation}
Note that in our notation $\mA_{\alpha\beta}
(\sigma,w)$ is a spatial component of
Lax connection.

Our goal is to calculate the 
Poisson bracket between $\mT(w)$ and
$\mT(v)$.  Following  \cite{Maillet:1985ek}
we consider the Poisson bracket
between  any dynamical
quantity $X_{\gamma\delta}$ and
$\mT_{\alpha\beta}(\sigma_1,\sigma_2,w)$ where 
$X_{\gamma\delta}$ does not depend  
$\sigma_1$ and $\sigma_2$
\begin{equation}\label{mTW}
\pb{\mT_{\alpha\beta}(\sigma_1,\sigma_2,w),
X_{\gamma\delta}}=W_{\alpha\gamma,
\beta\delta}(\sigma_1,\sigma_2,w)  \ , 
\end{equation}
where
\begin{equation}
\partial_{\sigma_1} X_{\gamma\delta}=0 \ ,
\quad  \partial_{\sigma_2} X_{\gamma\delta}=0 \ . 
 \end{equation}
If we  derive  
(\ref{mTW}) with respect $\sigma_1$ and
$\sigma_2$ and   use 
(\ref{diffdefT}) we obtain two
differential equations for $W(\sigma_1,\sigma_2)$
\begin{eqnarray}\label{partialxW}
\partial_{\sigma_1} W_{\alpha\gamma,
\beta\delta} (\sigma_1,\sigma_2,w)=
\mA_{\alpha\sigma}(\sigma_1,w)
W_{\sigma\gamma,\beta\delta}(\sigma_1,\sigma_2,w)+
\pb{\mA_{\alpha\sigma}(\sigma_1,w),
X_{\gamma\delta}}
\mT_{\sigma\beta}(\sigma_1,\sigma_2,w)
\nonumber \\
\end{eqnarray}
and 
\begin{eqnarray}\label{partialyW}
\partial_{\sigma_2} W_{\alpha\gamma,
\beta\delta}(\sigma_1,\sigma_2,w)=
-\mT_{\alpha\sigma}(\sigma_1,\sigma_2,w)
\pb{\mA_{\sigma\beta}(\sigma_2,w),
X_{\gamma\delta}}-
W_{\alpha\gamma,\sigma\delta}(\sigma_1,\sigma_2,w)
\mA_{\sigma\beta}(\sigma_2,w) \ . 
\nonumber \\
\end{eqnarray}
The equations (\ref{partialxW}) and
(\ref{partialyW}) have solution in the
form 
\begin{eqnarray}\label{W}
W_{\alpha\gamma,\beta\delta}
(\sigma_1,\sigma_2,w)=
\int_{\sigma_2}^{\sigma_1}
 d\sigma' \mT_{\alpha\sigma_1}(\sigma_1,\sigma',w)
\pb{\mA_{\sigma_1\sigma_2}(\sigma',w),
X_{\gamma\delta}}
\mT_{\sigma_2\beta}(\sigma',\sigma_2,w) \ .
\nonumber \\
\end{eqnarray}
Let us now presume that $X_{\gamma\delta}=
\mT_{\gamma\delta}(\sigma'_1,\sigma'_2,v)$ where
all $\sigma_1,\sigma_2,\sigma_1'$ 
and $\sigma'_2$ are distinct.
Then (\ref{mTW}) together
with  (\ref{W}) implies 
\begin{eqnarray}
\pb{\mT_{\alpha\beta}(\sigma_1,\sigma_2,w),
\mT_{\gamma\delta}(\sigma'_1,\sigma'_2,v)}&=&
\int_{\sigma_2}^{\sigma_1} d\sigma
\int_{\sigma_2'}^{\sigma_1'}
d\sigma' \mT_{\alpha\sigma_1}(\sigma_1,\sigma,w)
\mT_{\gamma\rho_1}(\sigma_1',\sigma',v)\times 
\nonumber \\
&\times & 
\pb{\mA_{\sigma_1\sigma_2}(\sigma,w),
\mA_{\rho_1\rho_2}(\sigma',v)}
\mT_{\sigma_2\beta}(\sigma,\sigma_2,w)
\mT_{\rho_2\delta}(\sigma',\sigma'_2,v) \ . 
\nonumber \\
\end{eqnarray} 
Let us now presume that the Poisson bracket
of spatial components of Lax 
connection $\mA(\sigma,w)$ and $\mA(\sigma',v)$,
where $w$ and $v$ are spectral 
parameters, takes the form 
\begin{eqnarray}\label{pbmA1ap}
\pb{\mA_{\alpha\beta}(\sigma,w),
\mA_{\gamma\delta}(\sigma',v)}=
\bA_{\alpha\gamma,\beta\delta}(\sigma,w,v)
\delta(\sigma-\sigma')+\nonumber \\
+\bB_{\alpha\gamma,\beta\delta}(\sigma,\sigma',w,v)
\partial_{\sigma'} \delta(\sigma-\sigma')+
\bC_{\alpha\gamma,\beta\delta}(\sigma,\sigma',w,v)
\partial_{\sigma}\delta(\sigma-\sigma') \ . 
\nonumber \\
\end{eqnarray}
Then an antisymmetry of Poisson bracket implies
\begin{eqnarray}\label{gendefbA}
\bA_{\alpha\gamma,\beta\delta}(\sigma,w,v)&=&
-\bA_{\gamma\alpha,\delta\beta}(\sigma,v,w) \ , 
\nonumber \\
\bB_{\alpha\gamma,\beta\delta}(\sigma,\sigma',w,v)&=&
-\bC_{\gamma\alpha,\beta\delta}(\sigma',\sigma,v,w)  \ , 
\nonumber \\
\bC_{\alpha\gamma,\beta\delta}(\sigma,\sigma',w,v)&=&
-\bB_{\gamma\alpha,\delta\beta}(\sigma',\sigma,v,w) \ . 
\nonumber \\
\end{eqnarray}
 Let us  introduce matrices
$r_{\alpha\gamma,\beta\delta}(\sigma,w,v),
s_{\alpha\gamma,\beta\delta}(\sigma,w,v)$ defined
as 
\begin{eqnarray}\label{rsdef}
s_{\alpha\gamma,\beta\delta}(\sigma,w,v)&=&
\frac{1}{2}[\bB_{\alpha\gamma,\beta\delta}
(\sigma,\sigma,w,v)+
\bB_{\gamma\alpha,\delta\beta}(\sigma,\sigma,w,v)]=
\nonumber \\
&=&\frac{1}{2}[\bB_{\alpha\gamma,\beta\delta}
(\sigma,\sigma,w,v)-
\bC_{\alpha\gamma,\beta\delta}(\sigma,\sigma,w,v)] \ , 
\nonumber \\
r_{\alpha\gamma,\beta\delta}(\sigma,w,v)&=&
\frac{1}{2}[\bB_{\alpha\gamma,\beta\delta}(
\sigma,\sigma,w,v)-\bB_{\gamma\alpha,\delta\beta}(\sigma,
\sigma,v,w)]+\hat{r}_{\alpha\gamma,\beta\delta}
(\sigma,w,v)=\nonumber \\
&=&\frac{1}{2}[\bB_{\alpha\gamma,\beta\delta}(
\sigma,\sigma,w,v)+\bC_{\alpha\gamma,
\beta\delta}(\sigma,
\sigma,v,w)]+\hat{r}_{\alpha\gamma,\beta\delta} 
(\sigma,w,v) \ , \nonumber \\
\end{eqnarray}
where $\hat{r}$ is solution of the
inhomogeneous first order differential equation
\begin{eqnarray}\label{hatr}
\partial_\sigma
\hat{r}_{\alpha\gamma,\beta\delta}
(\sigma,w,v)+[\hat{r}_{\alpha\gamma,\sigma\delta}(\sigma,w,v)
\mA_{\sigma\beta}(\sigma,w)-\mA_{\alpha\sigma}(\sigma,w)
\hat{r}_{\sigma\gamma,\beta\delta}(\sigma,w,v)]+
\nonumber \\
+[\hat{r}_{\alpha\gamma,\beta\sigma}(\sigma,w,v)
\mA_{\sigma\delta}(v,\sigma)-
\mA_{\gamma\sigma}(v,\sigma)\hat{r}_{\alpha\sigma,\beta\delta}
(\sigma,w,v)]=\Omega_{\alpha\gamma,\beta\delta}(\sigma,w,v) \ , 
\nonumber \\
\end{eqnarray}
where
\begin{eqnarray}
\Omega_{\alpha\gamma,\beta\delta}
(\sigma,w,v)&=&
\bA_{\alpha\gamma,\beta\delta}(\sigma,w,v)-
\partial_u (\bB_{\alpha\gamma,\beta\delta}
(\sigma,u,w,v)+\bC_{\alpha\gamma,\beta\delta}(u,\sigma,w,v))_{u=\sigma}+
\nonumber \\
&+&[\mA_{\gamma\sigma}(\sigma,v)\bB_{\alpha\sigma,\beta\delta}
(\sigma,\sigma,w,v)-\bB_{\alpha\gamma,\beta\sigma}(\sigma,\sigma,
w,v)\mA_{\sigma\delta}(\sigma,v)]+\nonumber \\
&+&[\mA_{\alpha\sigma}(\sigma,w)\bC_{\sigma\gamma,\beta\delta}
(\sigma,\sigma,w,v)-\bC_{\alpha\gamma,\sigma \delta}
(\sigma,\sigma,w,v)\mA_{\sigma\beta}(\sigma,w)] \ . 
\nonumber \\
\end{eqnarray}
Then we
 can rewrite the Poisson bracket
(\ref{pbmA1}) 
in the form
\begin{eqnarray}\label{pbMaillet}
\pb{\mA_{\alpha \beta}
(w,\sigma),\mA_{\gamma \delta}
(v,\sigma')}
=\left[r_{\alpha \gamma, \rho\delta}
(w,v,\sigma)\mA_{\rho\beta}(\sigma,w)
-\mA_{ \alpha \rho}(\sigma,w)r_{\rho 
\beta, \gamma\delta}(w,v,\sigma)+\right.
\nonumber \\
\left.+
r_{\alpha\gamma,\beta\sigma}(w,v,\sigma)
\mA_{\sigma\delta}(\sigma,v)-
\mA_{\gamma\sigma}(\sigma,v)r_{\alpha\sigma,
\beta\delta}(w,v,\sigma)+\right. \nonumber \\
+\left.
s_{\alpha \gamma, \rho\delta}
(w,v,\sigma)\mA_{\rho\beta}(\sigma,w)
-\mA_{ \alpha \rho}(\sigma,w)s_{\rho 
\beta, \gamma\delta}(w,v,\sigma)
-\right.\nonumber \\
\left.-
s_{\alpha\gamma,\beta\sigma}(w,v,\sigma)
\mA_{\sigma\delta}(\sigma,v)-
\mA_{\gamma\sigma}(\sigma,v)s_{\alpha\sigma,
\beta\delta}(w,v,\sigma)
\right]\delta(\sigma-\sigma')-
\nonumber \\
-(r(\sigma,w,v)+s(\sigma,w,v)-
r(\sigma',w,v)+s(\sigma',w,v))_{\alpha\gamma,\beta\delta}
\partial_\sigma\delta(\sigma-\sigma') \ . 
\nonumber \\
\end{eqnarray}
Let us now return to the equation
(\ref{hatr}). The general solution of
this equation takes the form 
\begin{eqnarray}
& &\hat{r}_{\alpha\gamma,\beta\delta}
(\sigma,w,v)= \nonumber \\
&=&\int_a^\sigma
d\sigma' \mT_{\alpha\sigma_1}(\sigma,\sigma',w)
\mT_{\gamma\sigma_1}(\sigma,\sigma',v)
\Omega_{\sigma_1\sigma_2,\rho_1\rho_2}(\sigma',w,v)
\mT_{\rho_1\beta}(\sigma',\sigma,w)\mT_{\rho_2\delta}
(\sigma',\sigma,v)+\nonumber \\
&+&\mT_{\alpha \sigma_1}
(\sigma,a,w)\mT_{\beta\sigma_2}
(\sigma,a,v)\hat{N}_{\sigma_1\sigma_2,
\rho_1\rho_2}(a,w,v)
\mT_{\rho_1\beta}(a,\sigma,w)
\mT_{\rho_2\gamma}(a,\sigma,v) \ , 
\nonumber \\
\end{eqnarray}
where $a$ is arbitrary real number and
$\hat{N}_{\alpha\beta,\gamma\delta}(w,v,a)$
is an arbitrary $\sigma$-independent
matrix that satisfy the relation
\begin{equation}
\hat{N}_{\gamma\alpha,\delta\beta}
(a,w,v)=-\hat{N}_{\alpha\gamma,\beta\delta}
(a,v,w) \ . 
\end{equation}
Note that the  $\hat{N}$ part of $\hat{r}$
is solution of the 
homogeneous equation associated with
(\ref{hatr}) and has to be determined by
choice of boundary conditions for
$\hat{r}(\sigma,w,v)$.
We would like also to mention that
$\hat{r}$ can be non-local expressions
in terms of canonical variables of theory.
Then, when it is possible, we can choose
$\hat{N}$ such that $\hat{r}$
be a local matrix in terms of field of the
theory.

Using of the form of the Poisson bracket
(\ref{pbMaillet}) we can calculate the
algebra of monodromy matrices when 
$\sigma_1,\sigma_2,\sigma'_1,
\sigma'_2$ are all different. 
We obtain, if $\sigma_1$ and $\sigma_1'$ 
are larger than $\sigma_2$ and $\sigma_2'$,
$\sigma_1^0=\mathrm{min}(\sigma_1,\sigma_1'),\quad
\sigma^0_2=\mathrm{max}(\sigma_2,\sigma_2')$ 
\begin{eqnarray}\label{algTT}
& &\pb{\mT_{\alpha\beta}(\sigma_1,\sigma_2,w),
\mT_{\gamma\delta}(\sigma_1',\sigma_2',v)}=
\nonumber \\
&=&\mT_{\alpha\sigma_1}(\sigma_1,\sigma_1^0,w)
\mT_{\gamma\sigma_2}(\sigma'_1,\sigma_1^0,v)
\left[r(\sigma_1^0,w,v)+\epsilon(\sigma_1-\sigma_1')
s(\sigma_1^0,w,v)\right]_{\sigma_1\sigma_2,\rho_1\rho_2}
\times \nonumber \\
&\times& \mT_{\rho_1\beta}(\sigma_1^0,\sigma_2,w)
\mT_{\rho_2\delta}(\sigma_1^0,\sigma'_2,v)-\nonumber \\
&-&\mT_{\alpha\sigma_1}(\sigma_1,\sigma_2^0,w)
\mT_{\gamma\sigma_2}(\sigma_1',\sigma^0_2,v)
\left[r(\sigma_2^0,w,v)+\epsilon(\sigma_2'-\sigma_2)
s(\sigma_2^0,w,v)
\right]_{\sigma_1\sigma_2,\rho_1\rho_2}
\times \nonumber \\
&\times& \mT_{\rho_1\beta}(\sigma_2^0,\sigma_2,w)
\mT_{\rho_2\delta}(\sigma_2^0,\sigma_2',v) \ , 
\nonumber \\
\end{eqnarray}
where $\epsilon(x)=\mathrm{sign}(x)$. 
It is important to note that in the
non-ultralocal case the algebra 
(\ref{algTT}), due to the presence of
the $s$-term, the function
\begin{equation}
\triangle^{(1)}_{\alpha\gamma,\beta\delta}
(\sigma_1,\sigma_2,
\sigma'_1,\sigma'_2,w,v)=
\pb{\mT_{\alpha\beta}(\sigma_1,\sigma_2,w),
\mT_{\gamma\delta}(\sigma'_1,\sigma'_2,v)}
\end{equation}
is well defined and continuous where
$\sigma_1,\sigma_2,\sigma_1',\sigma_2'$
are all distinct, but 
it has discontinuities
proportional to $2s$  across the hyperplanes
corresponding to some of the
$\sigma_1,\sigma_2,\sigma'_1,\sigma'_2$
being equal. 
Then if we want to define the Poisson
bracket of transfer matrices for coinciding
intervals ($\sigma_1=\sigma_1',
\sigma_2=\sigma'_2$) or adjacent 
intervals $(\sigma'_1=\sigma_2 \ 
\mathrm{or} \ \sigma_1=\sigma_1')$ requires
the value of the discontinuous
matrix-valued function $\triangle^{(1)}$
at its discontinuities.   It was shown
in \cite{Maillet:1985ec}
that requiring anti-symmetry of
the Poisson bracket and 
the derivation rule
to hold imposes the symmetric
definition of $\triangle^{(1)}$ at its
discontinuous points. For example,
at $\sigma_1=\sigma_1'$ we must define
\begin{eqnarray}\label{triangle}
\triangle^{(1)}_{\alpha\gamma,\beta\delta}
(\sigma_1,\sigma_2,\sigma_1,
\sigma'_2,w,v)&=&
\lim_{\epsilon\rightarrow 0^+}
\frac{1}{2}
(\triangle^{(1)}_{\alpha\gamma,\beta\delta}
(\sigma_1,\sigma_2,\sigma_1+
\epsilon,\sigma_2,w,v)+\nonumber \\
&+&\triangle^{(1)}_{\alpha\gamma,\beta\delta}
(\sigma_1,\sigma_2,
\sigma_1-\epsilon,\sigma'_2,w,v))
\nonumber \\
\end{eqnarray}
and likewise for all other possible
coinciding endpoints. This definition
of $\triangle^{(1)}$ at its discontinuities
implies an definition of the Poisson
bracket between transition matrices 
for coinciding and adjacent intervals
that is consistent with the anti-symmetry
 of the Poisson bracket and the derivation rule.  
However as was shown in 
\cite{Maillet:1985ek}
\footnote{For very nice recent discussion,
see \cite{Dorey:2006mx}.} this definition
of the Poisson bracket $\pb{\mT\stackrel{\otimes}{,}
\mT}$
\footnote{For simplicity of notation we use the double index
notation $\pb{\mT(w)\stackrel{\otimes}{,}
\mT(v)}_{\alpha\gamma,\beta\delta}=
\pb{\mT_{\alpha\beta}(w),\mT_{\gamma\delta}(v)}$.}
 does not satisfy the Jacobi 
identity so that in fact no strong definition
of the bracket 
$\pb{\mT\stackrel{\otimes}{,}
\mT}$ with coinciding or adjecent intervals
can be given without violating the
Jacobi identity. However, as was further shown in
\cite{Maillet:1985ek} it is possible to give
a weak definition of this bracket for coinciding
or adjacent intervals as well
\footnote{The bracket is called to be week
when the multiple bracket
$\pb{\mT\stackrel{\otimes}{,}\pb{\dots
\pb{\mT\stackrel{\otimes}{,}\mT}\dots}}$.
with $n$ factors of $\mT$ has to be defined for
each $n$ separately.}.
 We are not going
into details of the procedure, interesting
reader can read the original paper 
\cite{Maillet:1985ek} or more recent 
\cite{Dorey:2006mx}. For example, it was
shown that the algebra of two $\mT$'s
for equal intervals takes the form
\begin{eqnarray}\label{pbmT}
\pb{\mT_{\alpha\beta}
(\sigma_1,\sigma_2,w),\mT_{\gamma\delta}(\sigma_1,
\sigma_2,v)}&=&
r_{\alpha\gamma,\sigma\rho}
(\sigma_1,w,v)\mT_{\sigma\beta}
(\sigma_1,\sigma_2,w)\mT_{\rho\delta}
(\sigma_1,\sigma_2,v)-\nonumber \\
&-&\mT_{\alpha\sigma}(\sigma_1,\sigma_2,w)
\mT_{\gamma\rho}(\sigma_1,\sigma_2,v)
r_{\sigma\rho,\beta\delta}(\sigma_2,w,v) \ ,
\nonumber \\
\end{eqnarray}
where $\pb{,}$ stands for the weak brackets
defined in (\ref{triangle}).

Let us now return to our model. 
Since the reduced sigma model is defined
on the infinite line it is natural to introduce 
following object  
\begin{equation}
\Omega(w)=\mT(\infty,-\infty,w) \ . 
\end{equation}
Further, it is also natural to define 
\begin{equation}
r(w,v)\equiv \lim_{\sigma\rightarrow \infty}
r(w,v,\sigma)=\lim_{\sigma\rightarrow -\infty}
r(w,v,\sigma) \ .
\end{equation} 
For  example, this condition
clearly holds for  world-sheet fields
that  vanish at asymptotic
infinity. Then using (\ref{pbmT}) we finally obtain
\begin{eqnarray}
\pb{\tr \Omega(w),\tr \Omega(v)}=
\pb{\Omega_{\alpha\alpha}(w),
\Omega_{\beta\beta}(v)}=\nonumber \\
=r_{\alpha\beta,\sigma_1\sigma_2}(w,v)
\Omega_{\sigma_1\alpha}(w)\Omega_{\sigma_2\beta}(v)
-\Omega_{\alpha\sigma_1}(w)\Omega_{\beta\sigma_2}
(v)r_{\sigma_1\sigma_2,\alpha\beta}(w,v)=
\nonumber \\
=r_{\alpha\beta,\sigma_1\sigma_2}(w,v)
\Omega_{\sigma_1\alpha}(w)\Omega_{\sigma_2\beta}(v)
-r_{\alpha\beta,\sigma_1\sigma_2}(w,v)
\Omega_{\sigma_1\alpha}(w)\Omega_{\sigma_2\beta}(v)
=0 \ .
\nonumber \\
\end{eqnarray}
Since $\tr \Omega(w)$ is generator of local
conserved charges the result given above
implies that these conserved 
charges  are in involution with respect to
brackets (\ref{triangle}). This 
result implies  an classical integrability of
given theory. 

\section*{Acknowledgements}

This work  was supported in part by the Czech Ministry of
Education under Contract No. MSM
0021622409, by INFN, by the MIUR-COFIN
contract 2003-023852 and , by the EU
contracts MRTN-CT-2004-503369 and
MRTN-CT-2004-512194, by the INTAS
contract 03-516346 and by the NATO
grant PST.CLG.978785.

\end{document}